\documentclass{article}
\usepackage{emulateapj,pstricks,apjfonts}

\def\be{\begin{equation}}
\def\ee{\end{equation}}
\def\ba{\begin{eqnarray}}
\def\ea{\end{eqnarray}}

\begin{document}

%\submitted{Submitted to ApJ Letters}

\lefthead{Unbiased Estimate of 
 Dark Energy Density}
\righthead{Wang \& Lovelace}

\title{Unbiased Estimate of 
 Dark Energy Density from
 Type Ia Supernova Data}

\author{Yun Wang$^1$, \& Geoffrey Lovelace}

\affil{Department of Physics \& Astronomy\\
University of Oklahoma, Norman, OK 73019}

\footnotetext[1]{wang@mail.nhn.ou.edu}

\setcounter{footnote}{1}

\begin{abstract}

Type Ia supernovae (SNe Ia) are currently the best probes of the dark energy
in the universe. To constrain the nature of dark energy in a model-independent 
manner, we allow the density of dark energy, $\rho_X(z)$, to be an 
arbitrary function of redshift. 
Using simulated data from a space-based supernova pencil beam survey,
we find that by optimizing the number of parameters used to parametrize
the dimensionless dark energy density, $f(z)=\rho_X(z)/\rho_X(z=0)$, we 
can obtain an unbiased estimate of both $f(z)$ and the fractional matter 
density of the universe $\Omega_m$ (assuming a flat universe and 
that the weak energy condition is satisfied).
A plausible supernova pencil beam survey 
(with a square degree field of view and for an observational duration of 
one year) can yield about 2000 SNe Ia with $0\le z \le 2$ (\cite{Wang00a}). 
Such a survey in space would yield SN peak luminosities with a combined intrinsic and 
observational dispersion of ${\rm \sigma} (m_{int})=0.16$ mag. We find that
for such an idealized survey, $\Omega_m$ can be measured to 10\% accuracy, and the
dark energy density can be estimated to $\sim$ 20\% to 
$z \sim 1.5$, and $\sim$ 20-40\% to $z \sim 2$, depending on the time dependence of
the true dark energy density. Dark energy densities which vary more
slowly can be more accurately measured.
For the anticipated SNAP mission, 
$\Omega_m$ can be measured to 14\% accuracy,
and the dark energy density can be estimated to $\sim$ 20\% to 
$z \sim 1.2$. Our results suggest that SNAP may gain much 
sensitivity to the time-dependence of the dark energy density 
and $\Omega_m$ by devoting more observational time to the central 
pencil beam fields to obtain more SNe Ia at $z>1.2$.

We use both maximum likelihood analysis and Monte Carlo (when appropriate) to determine
the errors of estimated parameters. We find that Monte Carlo analysis gives 
a more accurate estimate of the dark energy density than the maximum likelihood 
analysis.

\end{abstract}

%\keywords{Cosmology}

\section{INTRODUCTION}

Supernova data suggest that most of the energy in the universe is
unknown to us (\cite{Garna98a,Riess98,Perl99}). Much 
theoretical effort has been devoted to
exploring the possible candidates for the dark energy 
(for example, see \cite{Peebles88,Frieman95,Caldwell98,Sahni00}),
and investigating the constraints on the nature of the dark energy
that can be derived from observational data 
(for example, see 
\cite{White98,Garna98b,Stein99,Efsta99,Huterer00,Podariu00,Waga00,Ng01,Podariu01,Weller01}).

The most straightforward and promising probe of dark energy is the
distance-redshift relation derived from observations of cosmological
standard candles. At present, type Ia supernovae (SNe Ia) are our best
candidates for cosmological standard candles 
(\cite{Phillips93,Riess95,Phillips99,Branch}).
Most workers have concentrated on constraining the equation of
state of the dark energy $w_X$ from SN data. However, without making
specific assumptions about $w_X$ (for example,
assuming it to be a constant), it is extremely hard to
constrain $w_X$ using SN data (\cite{MBS00,Barger}).
Recently, Wang and Garnavich (2001) [WG01] showed that 
it is easier to
extract constraints on the dark energy density $\rho_X(z)$, instead of
$w_X(z)$, from data. This is because
there are multiple integrals relating $w_X(z)$ to the luminosity distance
of SNe Ia, $d_L(z)$, which results in
a ``smearing'' that obscures the difference between different
$w_X(z)$ (\cite{MBS00}). On the other hand, $\rho_X(z)$ is related 
to the time derivative of the comoving distance to SNe Ia, $r'(z)$; 
hence it is less affected by the smearing effect. 
The advantage of measuring $\rho_X(z)$ over that of measuring
$w_X(z)$ is confirmed by Tegmark (2001).

WG01 gave a proof of principle of an
adaptive iteration technique for extracting the dimensionless
dark energy density $f(z)=\rho_X(z)/\rho_X(z=0)$ as
an arbitrary function from future SN data.
They found that feasible future SN data will allow us
to clearly differentiate dark energy with density that
changes with time from a cosmological constant $\Lambda$; 
however, estimates of $\Omega_m$ and $f(z)$ tend to
be significantly biased for dark energy densities that
vary substantially with time.

In this paper, we apply a significantly improved and 
optimized version of the adaptive iteration technique for
obtaining an unbiased estimate of the dark energy density 
from SN Ia data.

\section{Dark Energy Density}
\label{sec:analysis}

The total energy density of the universe is
\ba
\rho(z) & = & \rho_m^0(1+z)^3 +\rho_k^0(1+z)^2 +\rho_X^0\, f(z)\\
 & = & \rho_c^0 \left[\Omega_m (1+z)^3 +\Omega_k(1+z)^2 + \Omega_X \,f(z)
\right],\nonumber 
\ea
where the superscript ``0'' indicates present values, $f(z=0)=1$, and
$\Omega_k=1-\Omega_m-\Omega_X=-k/H_0^2$.
If the unknown energy is due to a cosmological constant $\Lambda$,
$f(z)=1$. Clearly, $f(z)=\rho_X(z)/\rho_X(z=0)$ is a very good probe of
the nature of the unknown energy.
Following WG01, we impose the
weak energy condition (the energy density is nonnegative
for any observer [\cite{Wald84}]), which implies that
$f'(z)\ge 0$ (see WG01).

The equation of state of the dark energy is
\be
w_X(z) \equiv \frac{p_X(z)}{\rho_X(z)}=\frac{1}{3}(1+z) \frac{f'(z)}{f(z)} -1.
\ee
A constant $w_X(z)$ corresponds to $f(z) \propto (1+z)^{\alpha}$,
where $\alpha$ is a constant. The values $\alpha=0$, $\alpha=3$, and $\alpha=4$ 
correspond to a cosmological constant, matter, and radiation
respectively. 
We parametrize $f(z)$ as 
\ba
\label{eq:f(z)}
&&f(z)=\left( \frac{z_i-z}{z_i-z_{i-1}} \right)\, f_{i-1}+
\left( \frac{z-z_{i-1}}{z_i-z_{i-1}} \right)\, f_i,
 \,\,\, z_{i-1} < z \leq z_i, \nonumber \\
&& z_0=0, \,\, z_{n}=z_{max}; \hskip 1cm f_0=1
\ea
where $f_i$ ($i=1,2,...,n-1)$ are independent variables
to be estimated from data. We let $f_n$ be either an
independent variable, or linearly extrapolated from
$f_{n-1}$ and $f_{n-2}$, whichever gives the smaller 
$\chi^2$ per degree of freedom for the same $n$.

The measured distance modulus for a SN Ia is
$\mu_0^{(l)}= \mu_p^{(l)}+\epsilon^{(l)}$,
where $\mu_p^{(l)}$ is the theoretical prediction
$\mu_p^{(l)}= 5\,\log\left( d_L(z_l)/\mbox{Mpc} \right)+25$,
and $\epsilon^{(l)}$ is the uncertainty in the measurement, including
observational errors and intrinsic scatter in the SN Ia absolute
magnitudes. 
Denoting all the parameters to be fitted as {\bf s},
we can estimate {\bf s} using a modified $\chi^2$ statistic, 
which results from integrating the probability density function
for parameters {\bf s}, $ p(\mbox{\bf s}) \propto \exp(- \chi^2/2)$,
over the Hubble constant $H_0$. We write (see WG01)
\be
\label{eq:chi2mod}
\tilde{\chi}^2 \equiv \chi_*^2 - \frac{C_1}{C_2} \left( C_1+ 
\frac{2}{5}\,\ln 10 \right),
\ee
where
\ba
\chi_*^2 &\equiv& \sum_l \frac{1}{\sigma_l^2} \left( \mu_{p}^{*(l)}-
\mu_{0}^{(l)} \right)^2, \\
C_1 &\equiv& \sum_l \frac{1}{\sigma_l^2} \left( \mu_{p}^{*(l)}-
\mu_{0}^{(l)} \right), \hskip 1cm
C_2 \equiv \sum_l \frac{1}{\sigma_l^2},\\
\mu_p^* &\equiv &\mu_p(h=h^*)=42.384-5\log h^*+ 5\log \left[H_0 r(1+z)\right].
\ea
We take $h^*=0.65$. Our results are independent of the choice of $h^*$.

%Future cosmic microwave background (CMB) space missions 
%MAP (\cite{Bennett97}) and Planck (\cite{DeZotti99}), together with
%the galaxy redshift surveys SDSS (\cite{Gunn99}) and 2df (\cite{Dalton00}),
%will give us exquisitely accurate measurements of the geometry of the
%universe and the matter density in the universe 
%(\cite{Eisen99,Mike99,Wang99a}). 
%SN data can provide the unique probe on the nature of dark energy
%by allowing us to measure how the dark energy density varies
%with time.

The current cosmic microwave background (CMB) anisotropy measurements seem to 
indicate that we live in a flat universe (\cite{deBernardis00,Balbi00}).
Cluster abundances strongly suggest a low matter density 
universe (\cite{neta95,Carlberg96,neta98}). 
$\Omega_m=0.3$ and $\Omega_{\Lambda}=0.7$
is the best fit model to current observational data.
We will use $\Omega_m=0.3$ and $\Omega_X=0.7$ for our simulated data
in the rest of this paper.

In order to compare with the results of WG01, 
we consider the same two hypothetical 
dark energy models, given by
\ba
\label{eq:f(z)1}
f_q(z)&=& \frac{ e^{1.5z} }{(1+z)^{1.5}}, \hskip 1cm w_q(z)= -1+0.5z \nonumber \\
f_k(z) &=& \exp[0.9(1-e^{-z})], \hskip 1cm w_k(z)= 0.3 (1+z) e^{-z} -1; 
\ea
$f_q(z)$ and $f_k(z)$ represent quintessence 
($dw_q/dz >0$) and k-essence ($dw_k/dz <0$) models respectively
(\cite{Caldwell98,Armenda00}) and they
satisfy the weak energy condition
$f'(z)\ge 0$. 

\section{Results}

A feasible SN pencil beam survey 
(with a square degree field of view and for an effective observational 
period of one year) can yield almost 2000 SNe Ia out to $z=2$ 
(\cite{Wang00a}).
Let us combine the data from the SN pencil beam survey 
with SN data at smaller redshifts, so that
there are a minimum of 50 SNe Ia per 0.1 redshift interval
at any redshift.
This yields a total of 1966 SNe Ia for quintessence and
1898 SNe Ia for the k-essence model,
up to $z=2$ and for $\Omega_m=0.3$, $\Omega_X=0.7$.
We assume intrinsic 
and observational dispersions which are Gaussian with zero mean and
a variance of 0.16 magnitudes (for a space-based survey).

Fig.~1 shows the likelihood function ${\cal L}(\Omega_m)\propto
e^{-\chi^2_{min}(\Omega_m)/2}$ for the quintessence
model $f_q(z)$ and the k-essence model $f_k(z)$.  
Note that $\chi^2_{min}(\Omega_m)$
is marginalized over $n$ independent parameters that
parametrize the dimensionless dark energy density $f(z)$, 
$f_i$ ($i=1,2,...,n$) (see Eq.(\ref{eq:f(z)})).
For the quintessence model $f_q(z)$, we allow $f_n$ to be an 
independent variable, and for the k-essence model $f_k(z)$, 
we linearly extrapolate $f_{n-2}$ 
and $f_{n-1}$ to obtain $f_n$ (see text after Eq.[\ref{eq:f(z)}]).
The curves (peak location from left to right) correspond to 
$n=3, 4, 5, 10$ and $n=4, 5, 6, 10$ for $f_q(z)$ and $f_k(z)$
respectively. 

%%%%%%%%%%%%%%%%%%%%%%%%%%%%%%%%%%%%%%%%%%%%%%%%%%%%%%%%%%%%%%%%%%%%%%%%%%

\pspicture(0,0.2)(9,12.4)

\rput[tl]{0}(-0.2,15.3){\epsfxsize=9.cm \epsfclipon
\epsffile{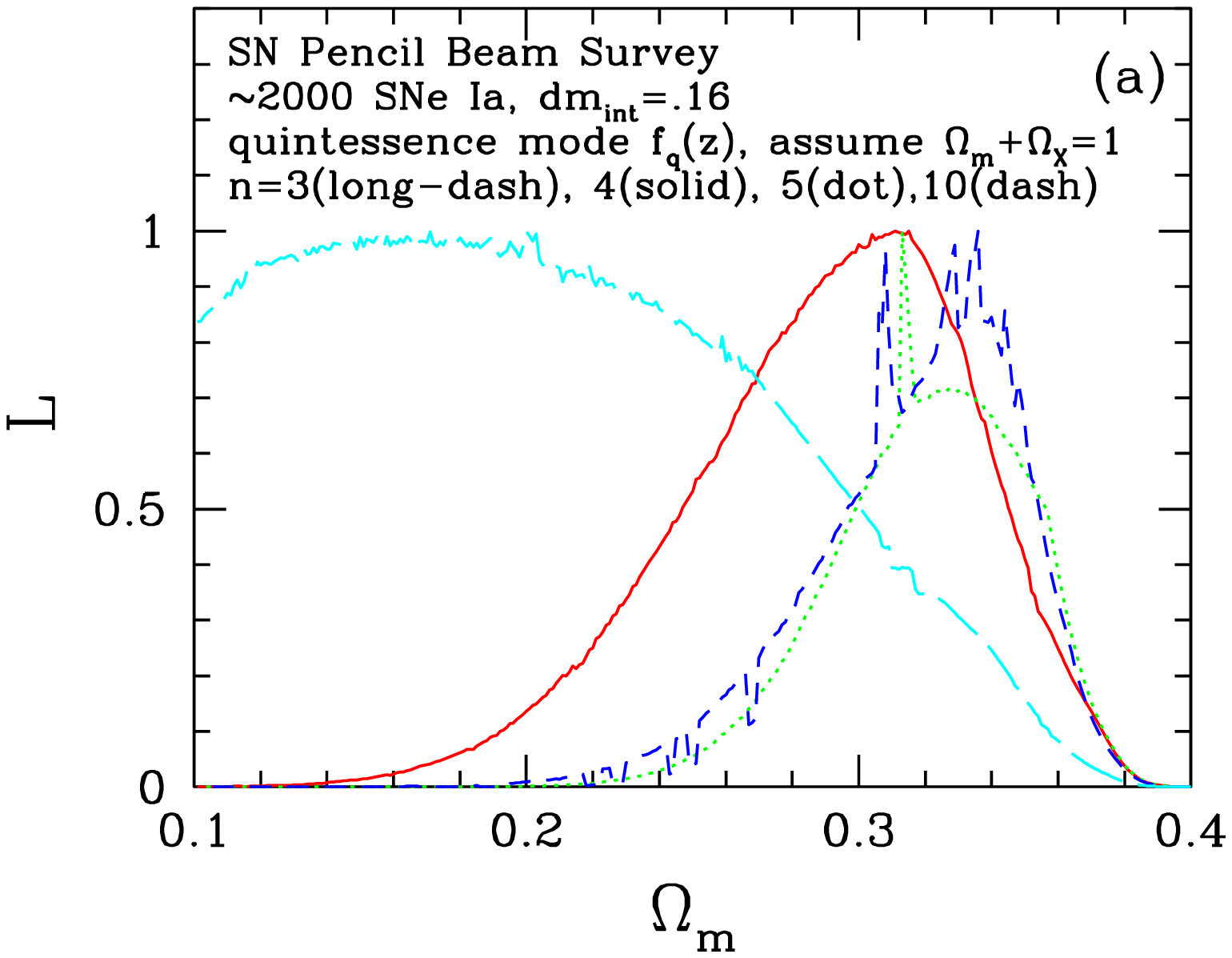}}

\rput[tl]{0}(-0.2,10.4){\epsfxsize=9.cm \epsfclipon
\epsffile{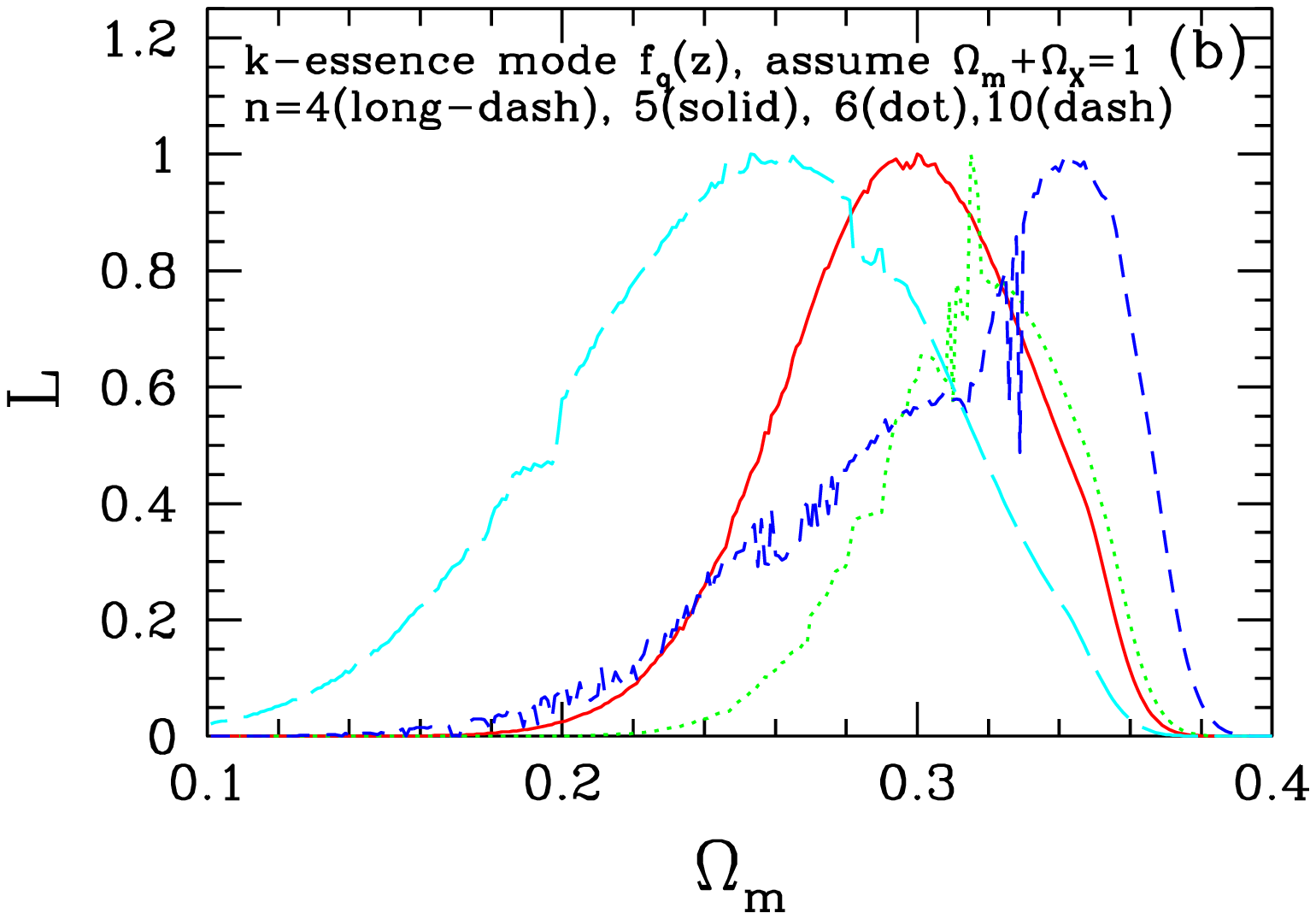}}

\rput[tl]{0}(0,2.2){
\begin{minipage}{8.75cm}
\small\parindent=3.5mm
{\sc Fig.}~1.---The likelihood function ${\cal L}(\Omega_m) \propto 
\exp\left[-\chi^2_{min}(\Omega_m)/2 \right]$ for the quintessence
model $f_q(z)$ and k-essence model $f_k(z)$. 
%
%\par
\end{minipage}
}
\endpspicture

%%%%%%%%%%%%%%%%%%%%%%%%%%%%%%%%%%%%%%%%%%%%%%%%%%%%%%%%%%%%%%%%%%%%%%%%%%
\vspace{-0.3in}

For the quintessence model $f_q(z)$,
the $\chi^2$ per degree of freedom decreases
as we decrease $n$ from $n=10$ to $n=4$, with a modest shift
in the most likely value of $\Omega_m$  (see Fig.1a); 
this is as expected
because our estimate of $\Omega_m$ should be roughly independent
of the parametrization of $f(z)$. There is a substantial shift 
in the most likely value of $\Omega_m$ as we change $n$ from
$n=4$ to $n=3$. Although the $\chi^2$ per degree of freedom
is smaller for $n=3$, the optimal choice is $n=4$ for this
model, since a greater degree of degeneracy between $\Omega_m$ 
and $f_i$ ($i=1,2,...,n$) sets in for $n=3$, which renders the 
estimated value of $\Omega_m$ significantly biased. For real data, 
and the true value of $\Omega_m$ unknown, this sudden increase in 
degeneracy can be inferred via the substantial shift in the 
estimated value of $\Omega_m$.

For the k-essence model $f_k(z)$, 
the $\chi^2$ per degree of freedom decreases
as we decrease $n$ from $n=10$ to $n=4$, with the largest shift
in the most likely value of $\Omega_m$ ocurring as we change $n$
from $n=5$ to $n=4$ (see Fig.1b). Hence $n=5$ is the optimal 
choice for this model.
Note that the transition that marks the sudden increase in degeneracy
(at $n=4$) is less dramatic than in the case of the quintessence model 
$f_q(z)$.

Fig.~2 shows the dimensionless dark energy density $f_q(z)$ and $f_k(z)$
estimated with $n=4$ and $n=5$ respectively (see Fig.~1). 
The dashed line and circles represent the estimate derived for an ideal
space-based supernova pencil beam survey (\cite{Wang00a}).
The dotted lines and triangles represent the estimate derived for 
the SNAP\footnote{see http://snap.lbl.gov} mission. 
The errors are (16\%, 84\%) intervals of the cumulative probability 
distribution of the estimated parameter for $10^3$ Monte Carlo samples.

%%%%%%%%%%%%%%%%%%%%%%%%%%%%%%%%%%%%%%%%%%%%%%%%%%%%%%%%%%%%%%%%%%%%%%%%%%
%\begin{figure*}[tb]
\pspicture(0,0.4)(9,14.1)
%\psgrid(0,0)(8.5,16)

\rput[tl]{0}(-0.2,15.6){\epsfxsize=9.cm \epsfclipon
\epsffile{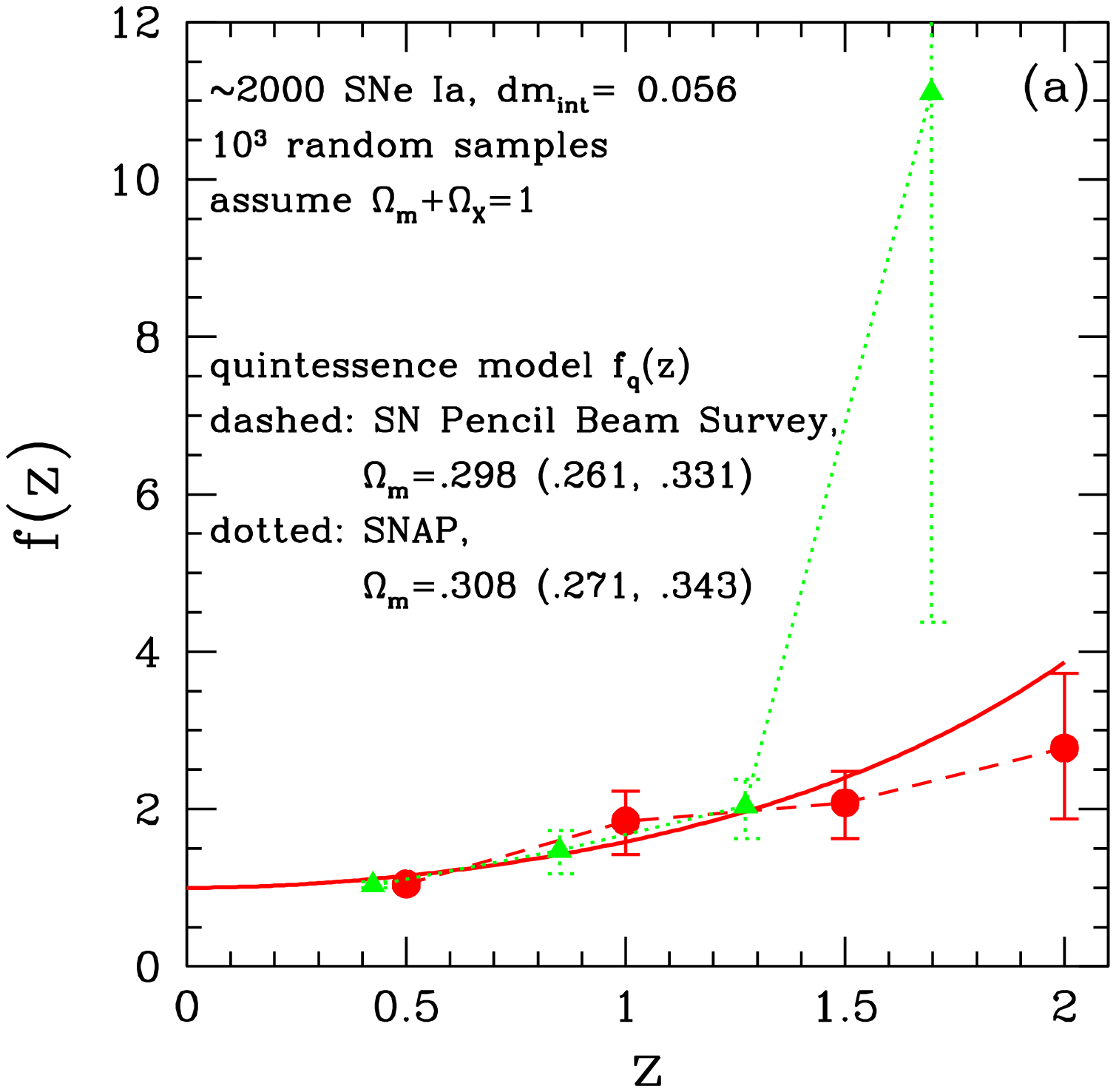}}

\rput[tl]{0}(-0.2,10.8){\epsfxsize=9.cm \epsfclipon
\epsffile{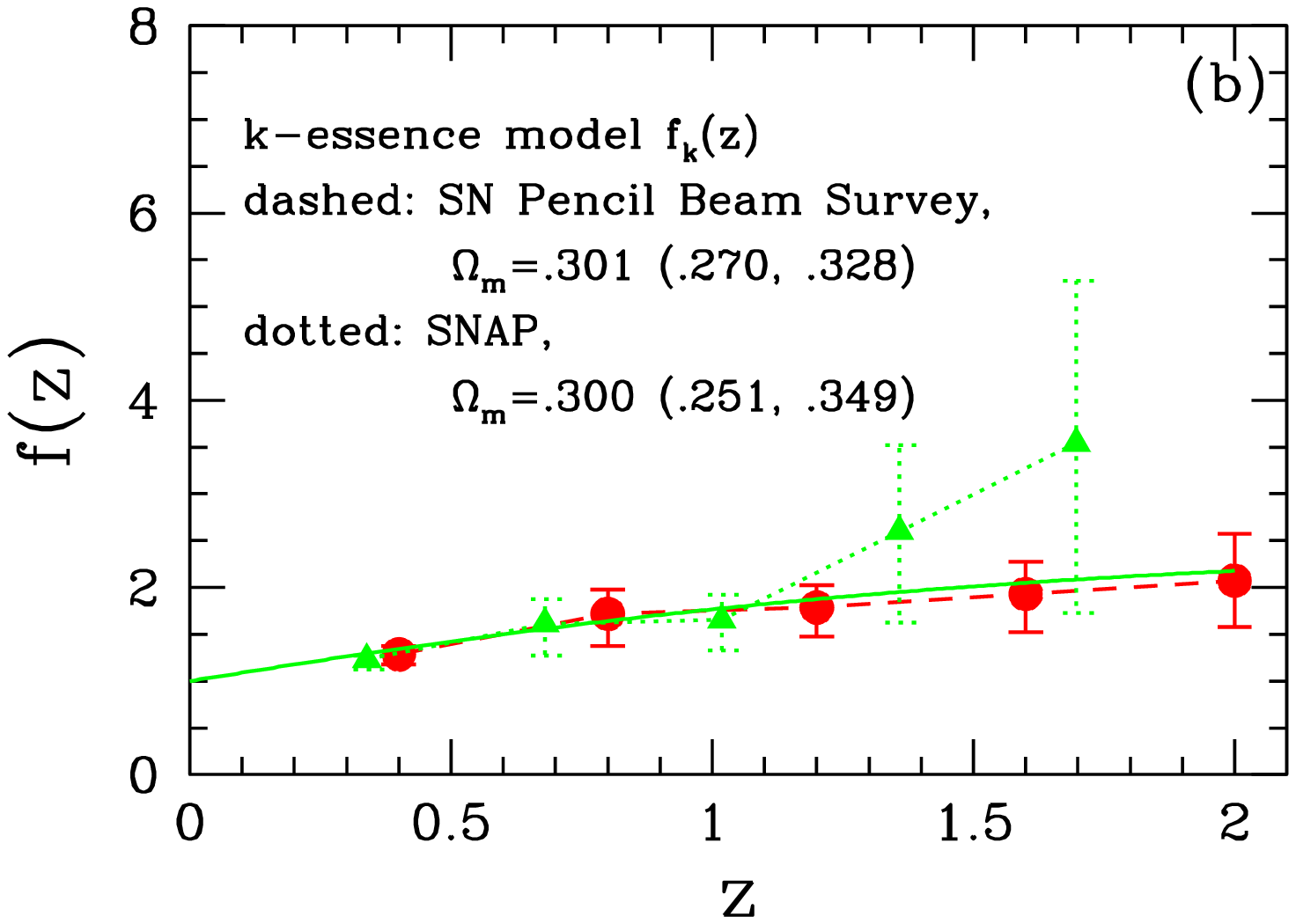}}

\rput[tl]{0}(0,2.2){
\begin{minipage}{8.75cm}
\small\parindent=3.5mm
{\sc Fig.}~2.---The dimensionless dark energy density $f_q(z)$ and $f_k(z)$
estimated with $n=4$ and $n=5$ respectively (see Fig.1). 
The dashed line and circles represent the estimate derived for an ideal
space-based supernova pencil beam survey (\cite{Wang00a}).
The dotted lines and triangles represent the estimate derived for 
the SNAP mission. 
\par
\end{minipage}
}
\endpspicture
%\end{figure*}
%%%%%%%%%%%%%%%%%%%%%%%%%%%%%%%%%%%%%%%%%%%%%%%%%%%%%%%%%%%%%%%%%%%%%%%%%%

Note that Fig.~1 and Fig.~2 differ significantly from
Fig.~2 of WG01, where the estimate 
of both $\Omega_m$ and $f_i$ ($i=1,2,...,n$) become significantly
more biased as $n$ is changed from $n=10$ to $n=6$. 
This work is greatly improved over that of WG01 
as follows:
(1) Maximum likelihood analysis is used here to determine the optimal
	choice of $n$ in parametrizing $f(z)$, which was not
	explored by WG01;
(2) The end point of $f(z)$, $f_n$, is allowed to differ from
	$f_{n-1}$ here, while $f_n=f_{n-1}$ was imposed in 
	WG01;
(3) The grid sizes of $\Delta \Omega_m=0.001$,
$\Delta f_i=0.05$ are used here, compared with the
grid sizes of $\Delta \Omega_m=0.02$, $\Delta f_i=0.1$ of
WG01;
(4) The errors of the estimated parameters are derived
via Monte Carlo from $10^3$ instead of $10^2$ random samples
as in WG01.

We have used a combination of maximum likelihood analysis and
Monte Carlo technique (when appropriate) in this paper. 
Fig.3 and Fig.4 show the probability distribution functions of $\Omega_m$ 
and $f_i$ ($i=1,2,3,4$) for the quintessence model $f_q(z)$.
The dotted curve is the likelihood function derived from
marginalized $\chi^2$, while the histogram is the probability 
distribution derived from Monte Carlo of $10^3$ random samples.
The arrows in each figure indicates the true value of the parameter.
Clearly, Monte Carlo analysis gives more accurate estimate of the 
dark energy density $f(z)$ than the maximum likelihood 
analysis for large $f(z)$.

%%%%%%%%%%%%%%%%%%%%%%%%%%%%%%%%%%%%%%%%%%%%%%%%%%%%%%%%%%%%%%%%%%%%%%%%%%

%\pspicture(0,0.2)(4,7.2)
\pspicture(0,0.2)(4,8.2)
\rput[tl]{0}(-0.2,10.6){\epsfxsize=9.cm \epsfclipon
\epsffile{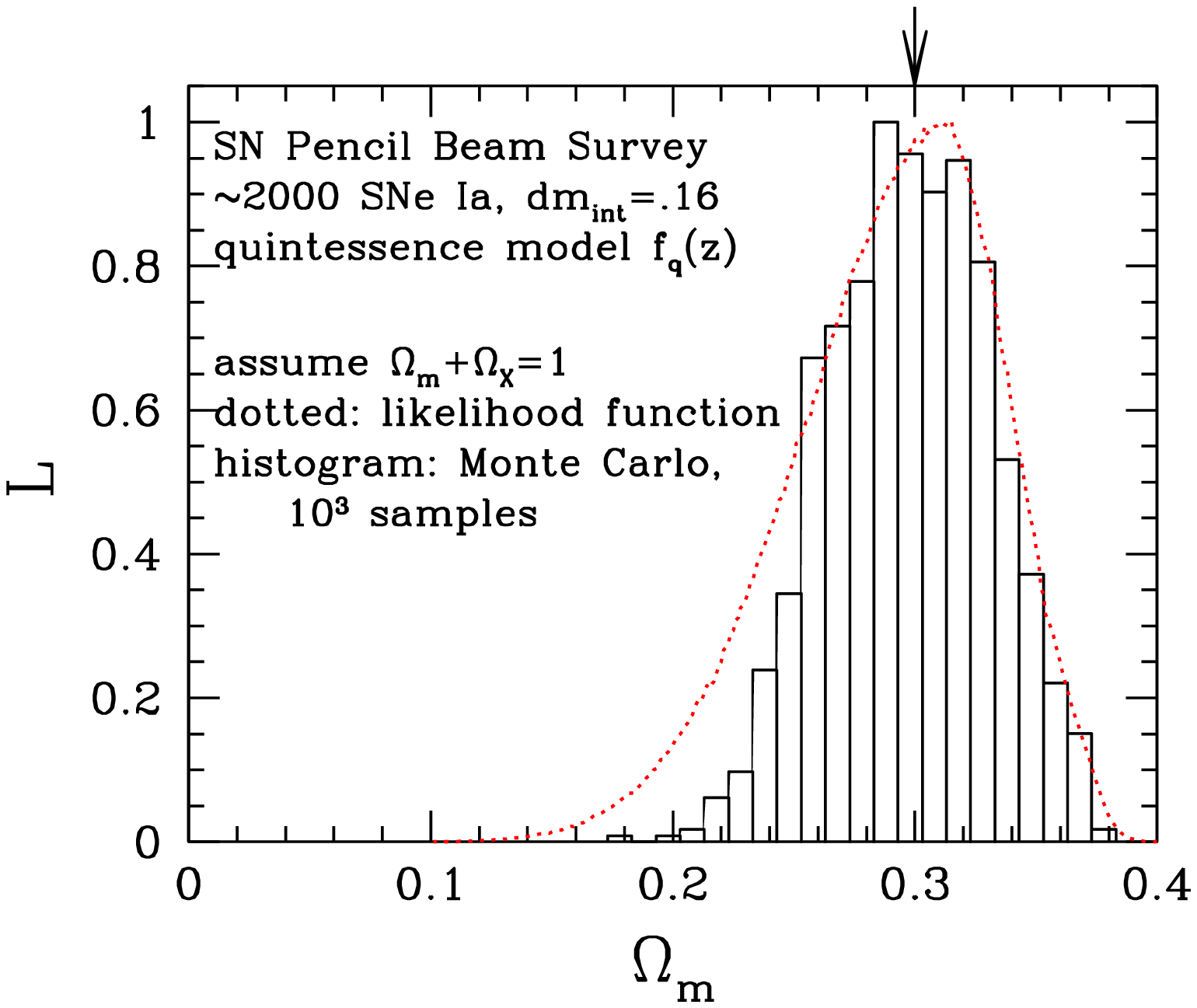}}

\rput[tl]{0}(0,2.2){

\begin{minipage}{8.75cm}
\small\parindent=3.5mm
{\sc Fig.}~3.---The probability distribution functions of $\Omega_m$ 
for the quintessence model $f_q(z)$.
The dotted curve is the likelihood function derived from
marginalized $\chi^2$, while the histogram is the probability 
distribution derived from Monte Carlo of $10^3$ random samples.
\par
\end{minipage}
}
\endpspicture

%%%%%%%%%%%%%%%%%%%%%%%%%%%%%%%%%%%%%%%%%%%%%%%%%%%%%%%%%%%%%%%%%%%%%%

%%%%%%%%%%%%%%%%%%%%%%%%%%%%%%%%%%%%%%%%%%%%%%%%%%%%%%%%%%%%%%%%%%%%%%%%%%
\begin{figure*}[tb]
%\pspicture(0,-0.3)(18.5,15.3)
\pspicture(0,-0.3)(18.5,13)
%\psgrid(0,-2)(18.5,16)

%\rput[tl]{0}(0.2,15.9){\epsfxsize=9.5cm \epsfclipon
\rput[tl]{0}(0.2,15.6){\epsfxsize=9.cm \epsfclipon
\epsffile{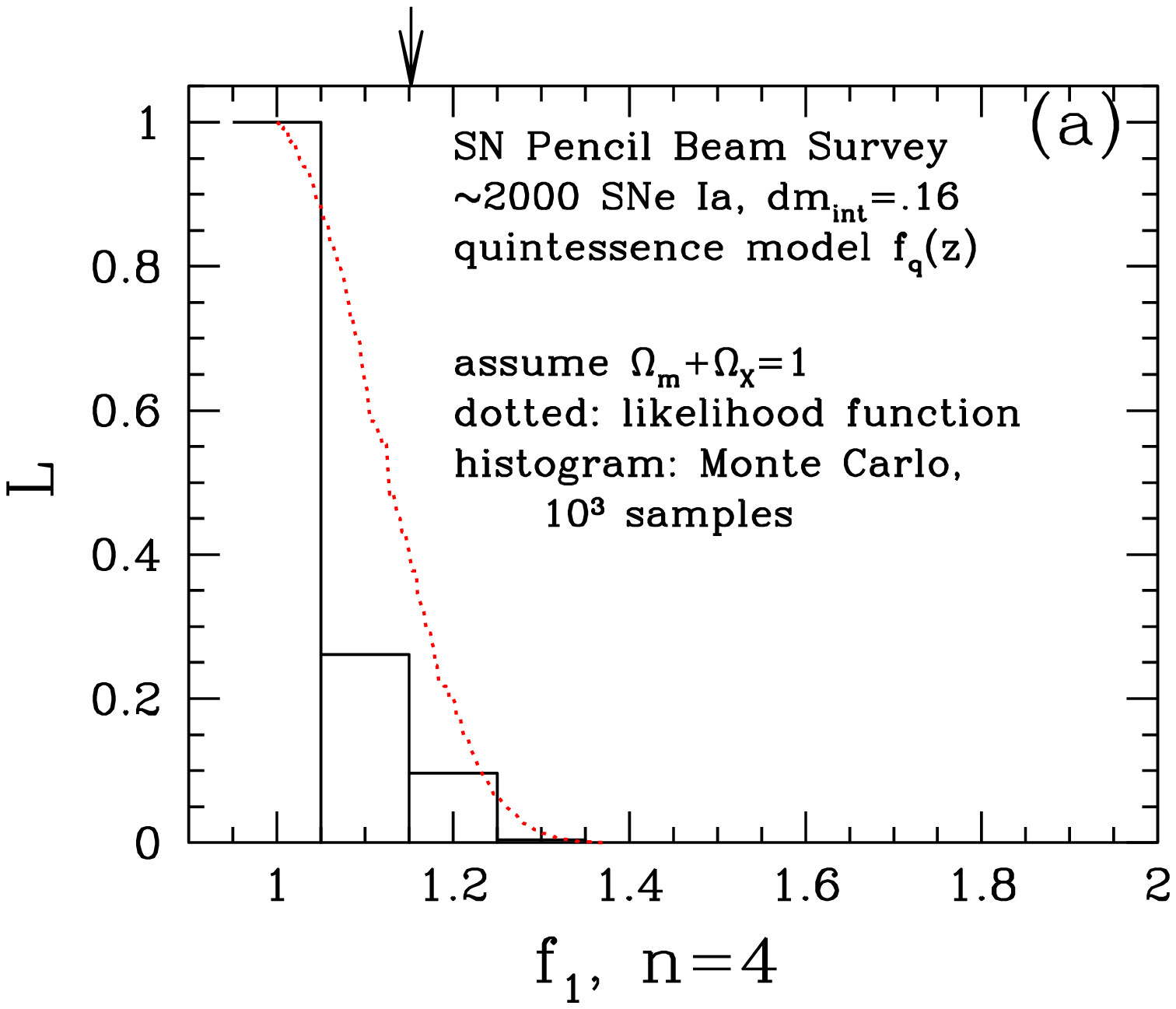}}

%\rput[tl]{0}(0.2,9.5){\epsfxsize=9.cm \epsfclipon
\rput[tl]{0}(0.2,9.8){\epsfxsize=9.cm \epsfclipon
\epsffile{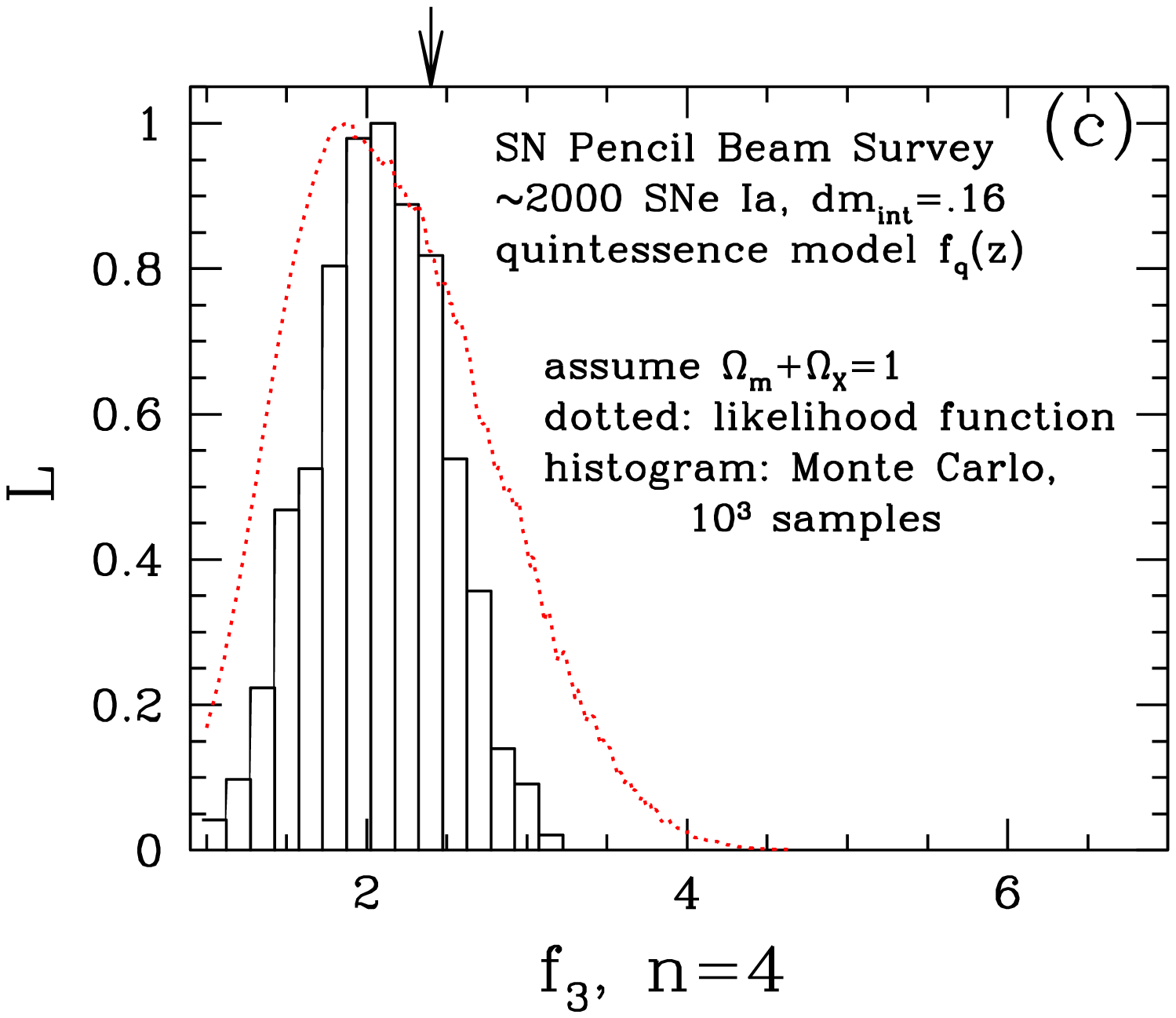}}

\rput[tl]{0}(9.4,15.6){\epsfxsize=9.cm \epsfclipon
\epsffile{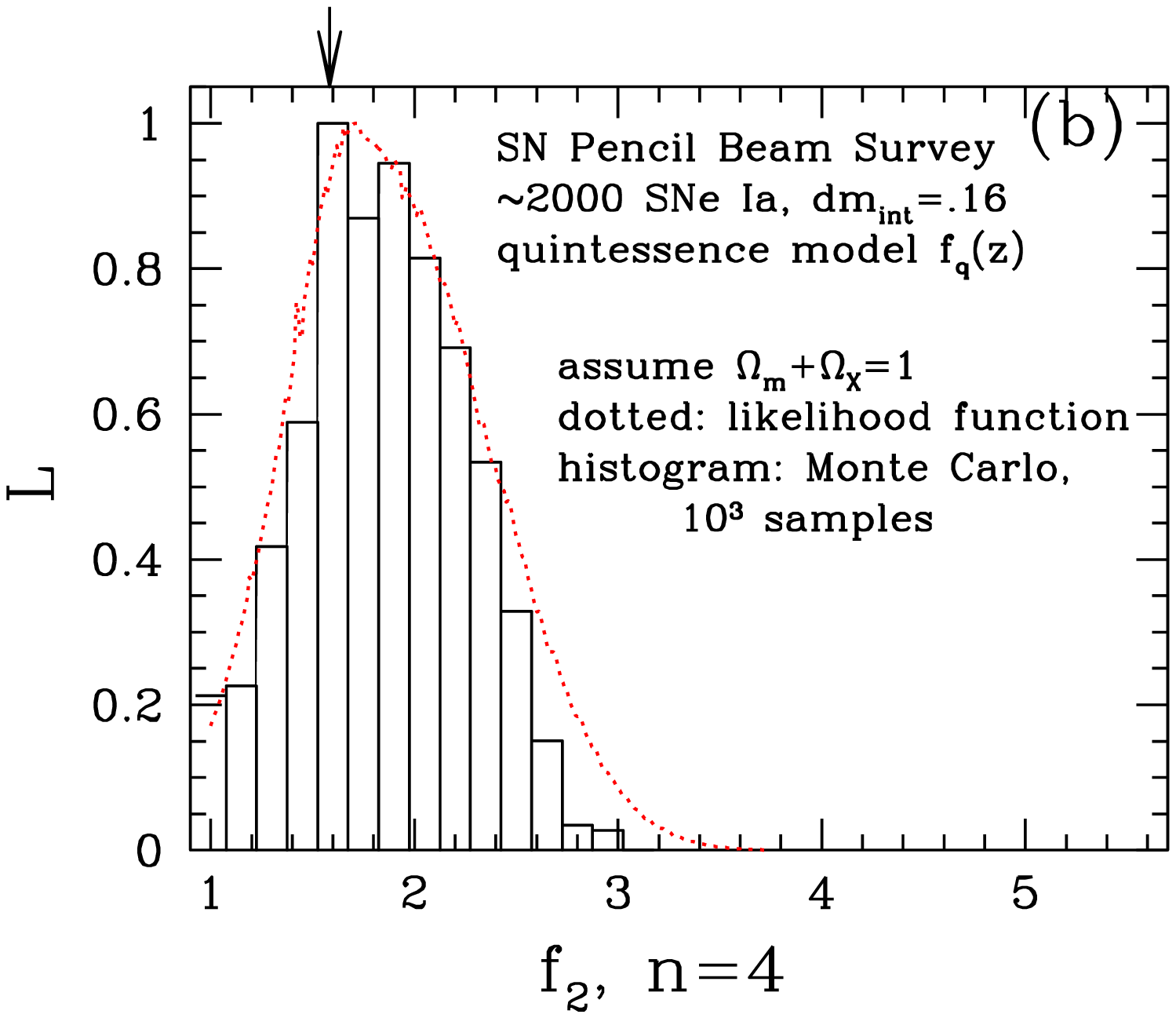}}

\rput[tl]{0}(9.4,9.8){\epsfxsize=9.cm \epsfclipon
\epsffile{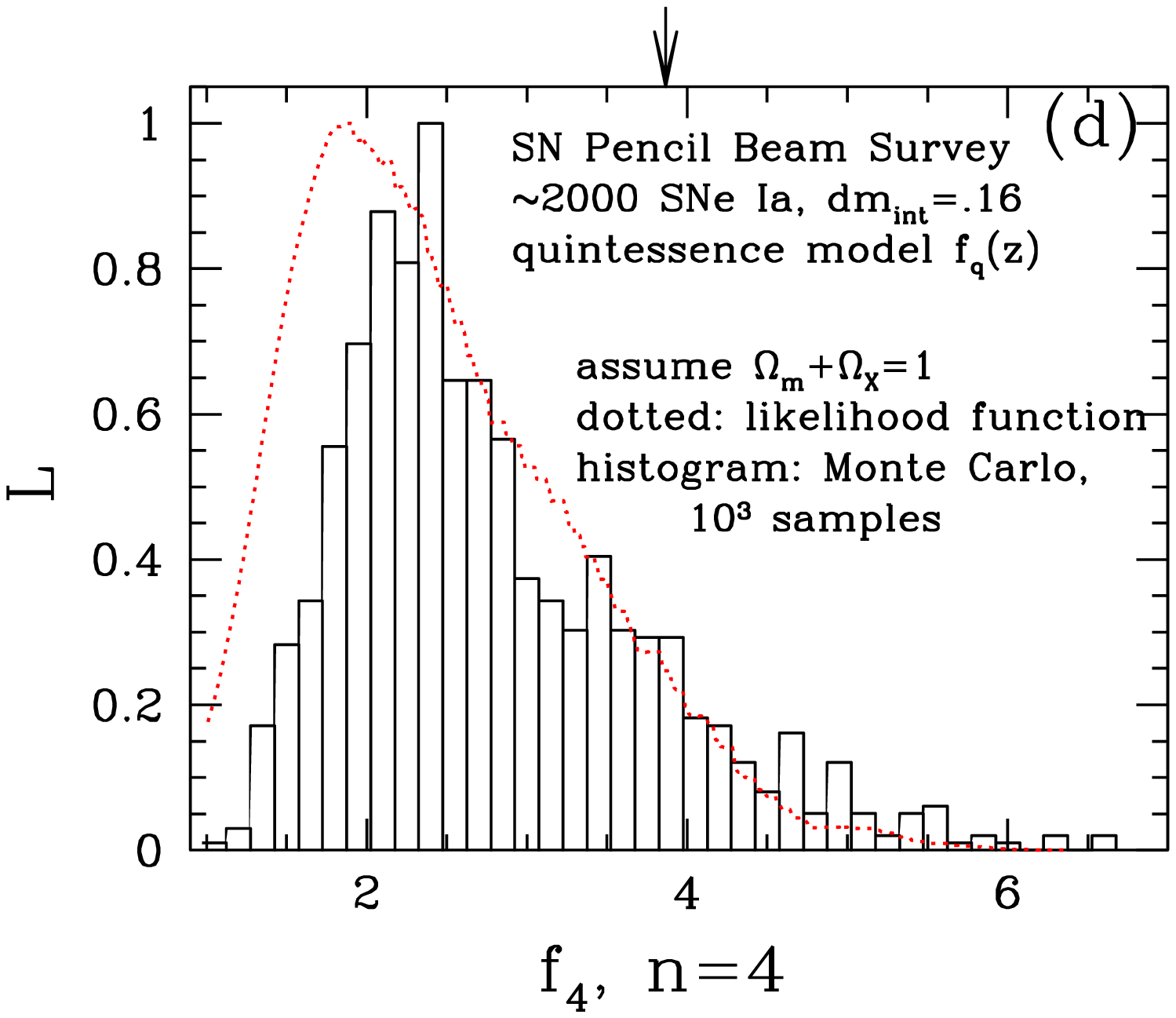}}

\rput[tl]{0}(-0.1,1.4){
\begin{minipage}{18.5cm}
\small\parindent=3.5mm
{\sc Fig.}~4.---The probability distribution functions of $f_i$ ($i=1,2,3,4$) for 
the quintessence model $f_q(z)$ with $n=4$.
The line types are the same as in Fig.~3.
\par
\end{minipage}
}
\endpspicture
\vspace{-0.3in}
\end{figure*}
%%%%%%%%%%%%%%%%%%%%%%%%%%%%%%%%%%%%%%%%%%%%%%%%%%%%%%%%%%%%%%%%%%%%%%%%%%
\vspace{-0.3in}

\section{Discussion}

Due to a substantial improvement in our adaptive iteration
technique (see WG01), we are able to explore much greater ranges 
of possibilities in the parametrization of the dimensionless dark 
energy density $f(z)=\rho_X(z)/\rho_X(z=0)$.
As a result, we have developed a method for finding unbiased estimates
of $\Omega_m$, together with $f(z)=\rho_X(z)/\rho_X(z=0)$ (as
an arbitrary function parametrized by its values at $n$ equally
spaced redshifts). The optimal choice of $n$ corresponds to
the smallest $\chi^2$ per degree of freedom without significant
shift in the estimated bestfit value of $\Omega_m$ (see Fig.1).
This leads to the unbiased estimate of $\Omega_m$, which is
crucial in deriving an unbiased estimate of the dimensionless 
dark energy density $f(z)$.

Our approach is more robust and model-independent than constraining 
the dark energy equation of state $w_X(z)$ (as a linear function of $z$)
assuming that $\Omega_m$ is known (\cite{Huterer00,Weller01}),
and complementary to the latter in probing the nature of the dark energy.

We find that for an ideal supernova pencil beam survey (\cite{Wang00a})
from space, $\Omega_m$ can be measured to 10\% accuracy, and 
$f(z)=\rho_X(z)/\rho_X(z=0)$ can be estimated to $\sim$ 20\% to 
$z \sim 1.5$, and 20-40\% to $z \sim 2$, depending on the time dependence of
the true $\rho_X(z)$ (see Fig.~2). Dark energy densities which 
vary more slowly can be more accurately measured, as might have
been expected, since it is easier in general to constrain slowly
varying functions compared with rapidly varying functions.

For the anticipated SNAP mission, $\Omega_m$ can be measured to 14\% accuracy,
and $f(z)=\rho_X(z)/\rho_X(z=0)$ can be estimated to $\sim$ 20\% to 
$z \sim 1.2$ (see Fig.~2). Compared with the idealized SN pencil beam survey 
(\cite{Wang00a}), 
the SNAP strategy is to obtain a larger number of SNe Ia (2408 versus
$~$2000 for a $1^{\circ}\times 1^{\circ}$ pencil beam) at $z\le 1.2$ by 
devoting a significant fraction of the observational
time on flanking $1^{\circ}\times 1^{\circ}$ fields that surround the
two central $1^{\circ}\times 1^{\circ}$ pencil beam fields, at the price 
of a sharply decreased number of SNe Ia at $z>1.2$;
this leads to large errors in estimated $f_i$ for $z>1.2$ (see
Fig.~2).
Our results suggest that SNAP may gain much
sensitivity to the time-dependence of $\rho_X(z)$ 
and $\Omega_m$ by devoting more 
observational time to the central pencil beam fields to obtain
more SNe Ia at $z>1.2$.

Future supernova pencil beam surveys hold
great promise for constraining the nature of dark energy 
in a model-independant manner.

\acknowledgements

We thank Eddie Baron and David Branch for helpful comments.
YW acknowledges support from NSF CAREER grant AST-0094335.

\end{document}